\documentclass[aps, prb, reprint, superscriptaddress, longbibliography, showkeys, citeautoscript, floatfix]{revtex4-2}

\usepackage[intlimits]{amsmath}
\usepackage{amssymb}
\usepackage{booktabs}
\usepackage{braket}
\usepackage{color}
\usepackage{float}
\usepackage[color=]{siunitx}
\usepackage{textgreek}
\usepackage{graphicx}
\usepackage{glossaries}
\usepackage{silence}
\usepackage[colorlinks=true,
            linkcolor=NavyBlue,
            citecolor=NavyBlue,
            filecolor=NavyBlue,
            urlcolor=NavyBlue
           ] {hyperref}
\usepackage[utf8]{inputenc}
\usepackage[version=4]{mhchem}
\usepackage{multirow}
\usepackage{soul} 
\usepackage{tabularx}
\usepackage[dvipsnames]{xcolor}
\usepackage[normalem]{ulem}
\usepackage{comment}
\usepackage{textcomp}

\usepackage{dcolumn}
\usepackage{float}
\usepackage{booktabs}
\usepackage{physics}

\newcommand{\rev}[1]{{\color{black}{#1}}}

\setacronymstyle{long-short}
\newacronym{tmd}{TMD}{transition metal dichalcogenide}
\newacronym{pdms}{PDMS}{polydimethylsiloxane}
\newacronym{vdw}{vdW}{van der Waals}
\newacronym{dft}{DFT}{density functional theory}
\newacronym{vis}{VIS}{visible}
\newacronym{nir}{NIR}{near-infrared}
\newacronym{uv}{UV}{ultraviolet}
\newacronym{mse}{MSE}{mean squared error}
\newacronym{si}{SI}{Supporting Information}
\newacronym{wse2}{WSe\textsubscript{2}}{Tungsted diselenide}

\newcommand{\dtu}{
    Department of Electrical and Photonics Engineering, Technical University of Denmark,
    2800, Kgs. Lyngby,
    Denmark
}

\newcommand{\innsbruck}{
    Institut für Experimentalphysik, Universität Innsbruck,
    6020 Innsbruck,
    Austria
}

\begin{document}

\title{High-purity and stable single-photon emission in bilayer \texorpdfstring{WSe\textsubscript{2}}{WSe2} via phonon-assisted excitation}

\author{Claudia Piccinini}
\affiliation{\dtu}

\author{Athanasios Paralikis}
\affiliation{\dtu}

\author{José Ferreira Neto}
\affiliation{\dtu}

\author{Abdulmalik A. Madigawa}
\affiliation{\dtu}

\author{Paweł Wyborski}
\affiliation{\dtu}

\author{Vikas Remesh}
\affiliation{\innsbruck}

\author{Luca Vannucci}
\affiliation{\dtu}

\author{Niels Gregersen}
\affiliation{\dtu}

\author{Battulga Munkhbat}
\email[]{bamunk@dtu.dk}
\affiliation{\dtu}

\begin{abstract}
\rev{
The excitation scheme is essential for single-photon sources, as it governs exciton preparation, decay dynamics, and the spectral diffusion of emitted photons. While phonon-assisted excitation has shown promise in other quantum emitter platforms, its proper implementation and systematic comparison with alternative excitation schemes have not yet been demonstrated in transition metal dichalcogenide (TMD) quantum emitters.
Here, we investigate the impact of various optical excitation strategies on the single-photon emission properties of bilayer WSe\textsubscript{2} quantum emitters. Based on our theoretical predictions for the exciton preparation fidelity, we compare excitation via the longitudinal acoustic and breathing phonon modes to conventional above-band and near-resonance excitations. Under acoustic phonon-assisted excitation, we achieve narrow single-photon emission with a reduced spectral diffusion of 0.0129 nm, a 1.8-fold improvement over above-band excitation. Additionally, excitation through breathing-phonon mode yields a high purity of $ 0.947\pm 0.079\,$ and reduces the decay time by over an order of magnitude, reaching $(1.33 \pm 0.04)\,$ns. Our comprehensive study demonstrates the crucial role of phonon-assisted excitation in optimizing the performance of WSe\textsubscript{2}-based quantum emitters, providing valuable insights for the development of single-photon sources for quantum photonics applications.
}
\end{abstract}

\maketitle
\section{Introduction}
\label{section:intro}
A key component in optical quantum information technologies is an on-demand source of single photons encoding the quantum information \cite{Heindel2023QuantumTechnology}. The fundamental requirement for a usable source is single photon emission with near-unity purity, efficiency, and indistinguishability.
Spontaneous parametric down-conversion \cite{Kwiat1995NewPairs} in a nonlinear material is a well-established method to generate highly indistinguishable photons \cite{Zhong201812-PhotonDown-Conversion}. However, its probabilistic nature limits the efficiency to merely a few percent. On the other hand, a two-level system \cite{Aharonovich2016Solid-stateEmitters} allows for the deterministic emission of single photons through the spontaneous emission process. Recently, highly efficient emission of indistinguishable photons has been achieved from semiconductor quantum dots (QDs) \cite{Wang2019TowardsMicrocavities, Ollivier2020ReproducibilitySources, Tomm2021APhotons} placed in carefully engineered microstructures. Nevertheless, epitaxial growth of high-quality QD material requires advanced and costly epitaxy tools.
As an alternative, quantum emitters (QEs) in two-dimensional (2D) transition metal dichalcogenides (TMDs) materials have recently emerged as an attractive platform for deterministic single-photon emission \cite{Srivastava2015Optically2, Tonndorf2015Single-photonSemiconductor, He2015SingleSemiconductors, Koperski2015SingleStructures, Chakraborty2015Voltage-controlledSemiconductor, R-PMontblanch2023LayeredTechnologies}. This platform provides several advantages, including accessible cost-effective materials, deterministic fabrication methods \cite{Gao2020RecentDichalcogenides, Parto2021DefectK, Micevic2022On-demandBeam}, and ease of integration with photonic \cite{Tonndorf2017On-ChipSource} and plasmonic systems \cite{Cai2017CouplingPolaritons, Blauth2018CouplingWaveguides}. Additionally, the reduced dimensionality of TMDs facilitates efficient light extraction \cite{Brotons-Gisbert2018EngineeringRegimes}. Consequently, single-photon emission over a broad spectrum from visible \cite{Barthelmi2020Atomistic} to telecommunication ranges \cite{Zhao2021Site-controlledMoTe2} has been demonstrated from mono- and few-layer TMDs via strain- and defect engineering \cite{Parto2021DefectK, So2021ElectricallyHeterostructure, Gao2023Atomically-thinCommunication, Iff2019Strain-TunableMonolayers, Linhart2019Localized2}.

Despite significant efforts in developing high-quality quantum emitters in TMDs, the efficient generation of indistinguishable single photons from these platforms remains an open fundamental challenge. Spectral diffusion and phonon-induced dephasing are the primary obstacles that degrade the figures of merit for single-photon sources. Strategies to mitigate these effects include integrating emitters into high-Q resonant cavities, stabilizing the charge environment, and implementing advanced optical excitation schemes. Recently, a strategy of coupling such quantum emitters to a resonant open cavity has been successfully demonstrated, exhibiting state-of-the-art efficiency as high as 65\% to the first lens, but limited indistinguishability of around 2\% \cite{Drawer2023Monolayer-BasedCoherence}. 
The same strategy was later employed to engineer the coherence of the emitted photons by selectively boosting the emission of the zero-phonon line \cite{Mitryakhin2024Electrodynamics}. Additionally, the spectral diffusion of the emitted photons has been improved via hBN encapsulation \cite{Parto2021DefectK, Daveau2020SpectralWrinkles}, substrate choice engineering \cite{Iff2017SubstrateEngineeringArchitectures} as well as with the implementation of electrical contacts \cite{Akbari2022Lifetime-LimitedModulation}.
\rev{
The choice of excitation scheme also significantly impacts the figures of merit for single-photon sources~\cite{Reindl2019HighlyDots}.
While continuous-wave (CW) resonant and non-resonant excitations of TMD quantum emitters have been reported 
\cite{Kumar2016ResonantWSe_2, Errando-Herranz2021ResonanceEmitters}, on-demand generation of single-photon generation requires pulsed excitation with short laser pulses.
}
Above-band excitation, though widely used, introduces photo-induced charge carriers and complex exciton dynamics through higher-order states that limit photon indistinguishability \cite{Flagg2010InterferenceDots}. In contrast, resonant excitation achieves near-unity indistinguishability \cite{He2013On-demandIndistinguishability}, but it is hindered by the reduced collection efficiency (maximum 50\%) inherent to the cross-polarization setup. 
\rev{Pulsed} phonon-assisted excitation offers a promising alternative, balancing efficient population inversion and high photon indistinguishability, previously demonstrated on semiconductor QDs~\cite{Thomas2021BrightDipole, Maring2024AVersatilePlatform}. This method leverages phonon scattering to mediate rapid relaxation to the exciton state, enabling efficient photon collection above 50\% while rejecting laser background. The feasibility of phonon-assisted excitation in TMD quantum emitters has been theoretically predicted~\cite{Vannucci2024Single-photonIndistinguishability}, yet its experimental implementation remains unexplored.


\begin{figure*}[hbt!]
\includegraphics[width=1 \textwidth]{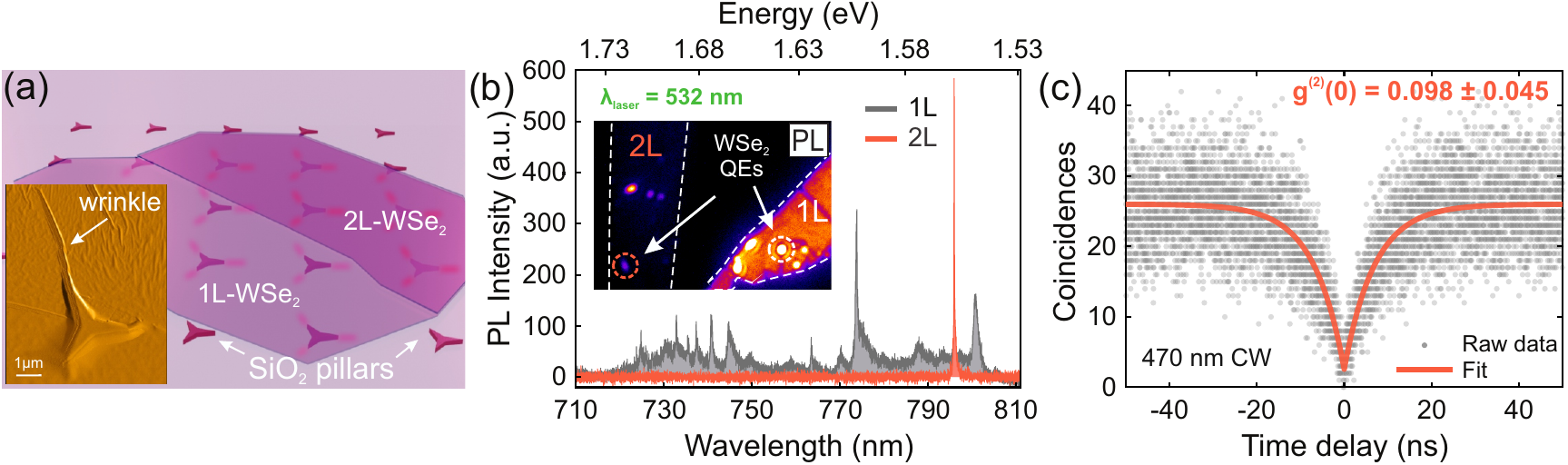}
\caption{\textbf{Quantum emitters in mono- and bilayer WSe\textsubscript{2}.} \textbf{(a)} Sketch of mono- (1L) and bilayer (2L) flake deposited on top of star-shaped nanostructures leading to the formation of nanowrinkles nearby their vertex. The nanowrinkles host the quantum emitters. Inset: atomic force microscopy (AFM) image of a star-shaped nanostructure covered by a bilayer flake with the visible presence of a nanowrinkle originating from the top vertex.
\textbf{(b)} \textmu PL spectra of the two bright spots in 1L WSe\textsubscript{2} (grey) and 2L WSe\textsubscript{2} (red) encircled in the inset image, integrated over \SI{1}{\second}. They show typical emission imprints of the two different thicknesses' flakes. Inset: color-coded PL image of the sample taken at $T$ = \SIlist{4}{\kelvin}. The white dashed lines highlight the contours of the 1L and 2L flakes. \textbf{(c)} Second-order correlation measurement of emitter Q1 from the bilayer spot (red dashed circle in inset in \textbf{(b)}) under CW LED excitation at 470 nm. The fit reveals a $g^{(2)}(0)$ value of $0.098 \pm 0.045$.}
\label{Figure1}
\end{figure*}

In this work, we systematically implement and compare three \rev{pulsed} optical excitation schemes, i.e., above-band, near-resonant, and phonon-assisted, to investigate their impacts on the single-photon properties of bilayer WSe\textsubscript{2} quantum emitters. We demonstrate high single-photon purity under pulsed excitation, reaching 0.921$\,\pm\,$0.015  and 0.943$\,\pm\,$0.018 for above-band and near-resonant excitation, respectively. Near-resonant excitation also yields shorter and simpler decay dynamics, with a decay constant of (12.62$\,\pm\,$1.18) ns.
\rev{
To explore phonon-assisted excitation, we first develop a theoretical model that describes a quantum emitter coupled to the phonon environment specific to bilayer WSe$_2$. Here, we consider both LA phonon modes, which are present in both mono- and bilayer WSe$_2$, and interlayer shear modes (SMs) and breathing mode (BM), which are unique to bilayer structures and exhibit energy dependence on the interlayer distance. Due to the inherent strong coupling to LA phonons, our theoretical calculations predict an efficient exciton population above 80\% under LA phonon-assisted excitation at a detuning of \SI{-0.4}{\nano\meter}. We also predict a sizable population at a larger detuning of \SI{-2.7}{\nano\meter}, dependent on the strength of the exciton-BM coupling.
Guided by our theoretical predictions, we experimentally demonstrate LA phonon-assisted excitation in WSe$_2$, achieving a two-fold reduction in spectral diffusion of the emission line compared to above-band excitation. Finally, we perform phonon-assisted excitation through the interlayer breathing mode. Owing to the larger detuning between the laser and the emitter's energy, we achieve full residual laser rejection and carry out single-photon characterization. We obtain a purity of 0.947$\,\pm\,$0.079 and a much faster decay rate ($\tau_1$ = (1.33$\,\pm\,$0.04) ns) in contrast to above-band and near-resonance excitations.
}
Our work contributes to the fundamental understanding of WSe\textsubscript{2} quantum emitter platform and to its application in future quantum photonics.

\section{System under study: mono- and bilayer WSe$_2$ quantum emitters}
\label{section:system}
We begin our study by preparing highly-polarized single-photon emitters out of WSe\textsubscript{2} flakes through \rev{a combination of} deterministic strain \rev{and defect} engineering as illustrated in Fig.~\ref{Figure1}(a). \rev{Based on the process presented in our previous work in Ref.~\cite{Paralikis2024TailoringEngineering}}, mechanically exfoliated mono- and bilayer WSe\textsubscript{2} flakes were transferred onto a thermally oxidized silicon substrate (d$_{\mathrm{SiO_2}}$ = \SI{110}{\nano\meter}) featuring an array of star-shaped nanostructures \rev{(d$_{\mathrm{pillars}}$ = \SI{150}{\nano\meter}). The long and sharp tips of the three-pointed star geometries induce strain on the material, generating spatially isolated nanowrinkles. Moreover, this type of geometry can host multiple nanowrinkles per site, increasing the number of available quantum emitters. The localized one-dimensional strain alters the energy bands leading to the efficient migration of free charge carriers to the area \cite{R-PMontblanch2023LayeredTechnologies}, while the orientation of the strain dictates the directionality of the linear polarization. Subsequently, deterministically introducing defects to the sample via e-beam irradiation has previously been shown to increase the quantum emitter yield in this platform \cite{Paralikis2024TailoringEngineering, Parto2021DefectK}. The \textit{Methods} section (\ref{section:methods}) provides further details on the fabrication process.}
To investigate the optical properties of the fabricated samples, we mounted them in an optical cryostat operating at \SI{4}{\kelvin}. The cryostat is equipped with nanopositioners and a low-temperature objective lens (NA = 0.82, $60\times$). Initially, a low-temperature photoluminescence (PL) image of the sample was acquired under \SI{470}{\nano\meter} LED excitation. A \SI{700}{\nano\meter} long-pass optical filter was employed to eliminate reflected signals from the excitation. As depicted in the inset of Fig.~\ref{Figure1}(b), the PL image reveals bright spots originating from the \rev{nanowrinkles in} mono- and bilayer WSe\textsubscript{2}, which are distinctly contrasted against the emission signal from the planar, unstrained flakes.

For a quantitative comparison between the mono and bilayer emitters, we measured the PL spectra of the individual bright spot regions of each under a 532 nm pulsed excitation (pulse duration \SI{200}{\femto\second}, repetition rate \SI{80}{\mega\hertz}) using a spectrometer ($f = 550$ mm, \SI{1200}{ grooves\per\milli\meter} grating). Fig.~\ref {Figure1}(b) shows the emission spectra obtained from two exemplary mono- and bilayer regions (dashed red and white circles in the inset) hosting strain-engineered quantum emitters. The monolayer emission spectrum exhibits numerous intense and narrow emission lines spreading across a broad spectral range of 720--\SI{810}{\nano\meter} on top of a broad, weak emission background from the planar monolayer. The observed narrow lines are attributed to single-photon emission from several different emitters \cite{Srivastava2015Optically2, Chakraborty2015Voltage-controlledSemiconductor, Tonndorf2015Single-photonSemiconductor}, while the broad background is ascribed to the emission from the planar monolayer region around the emitter, present in the measurement due to the diffraction-limited collection spot. On the other hand, the bilayer quantum emitter presents a single isolated narrow emission line (FWHM = \SI{0.14}{\nano\meter}) without any noticeable background, \rev{due to the indirect nature of the bandgap in pristine bilayer \cite{Zhao2013EvolutionWse2}}.
Surprisingly, despite the bilayer WSe\textsubscript{2} host being an indirect bandgap, the intensities of the emission lines are comparable to those observed in the monolayer under identical experimental conditions (more exemplary spectra in Fig. S1 in the Supplementary Information).
\rev{These considerations make WSe\textsubscript{2} bilayer a preferable platform over monolayer for our scopes. We will focus on bilayer QEs in the following.}


\begin{figure*}[hbt!]
\includegraphics[width=0.95 \textwidth]{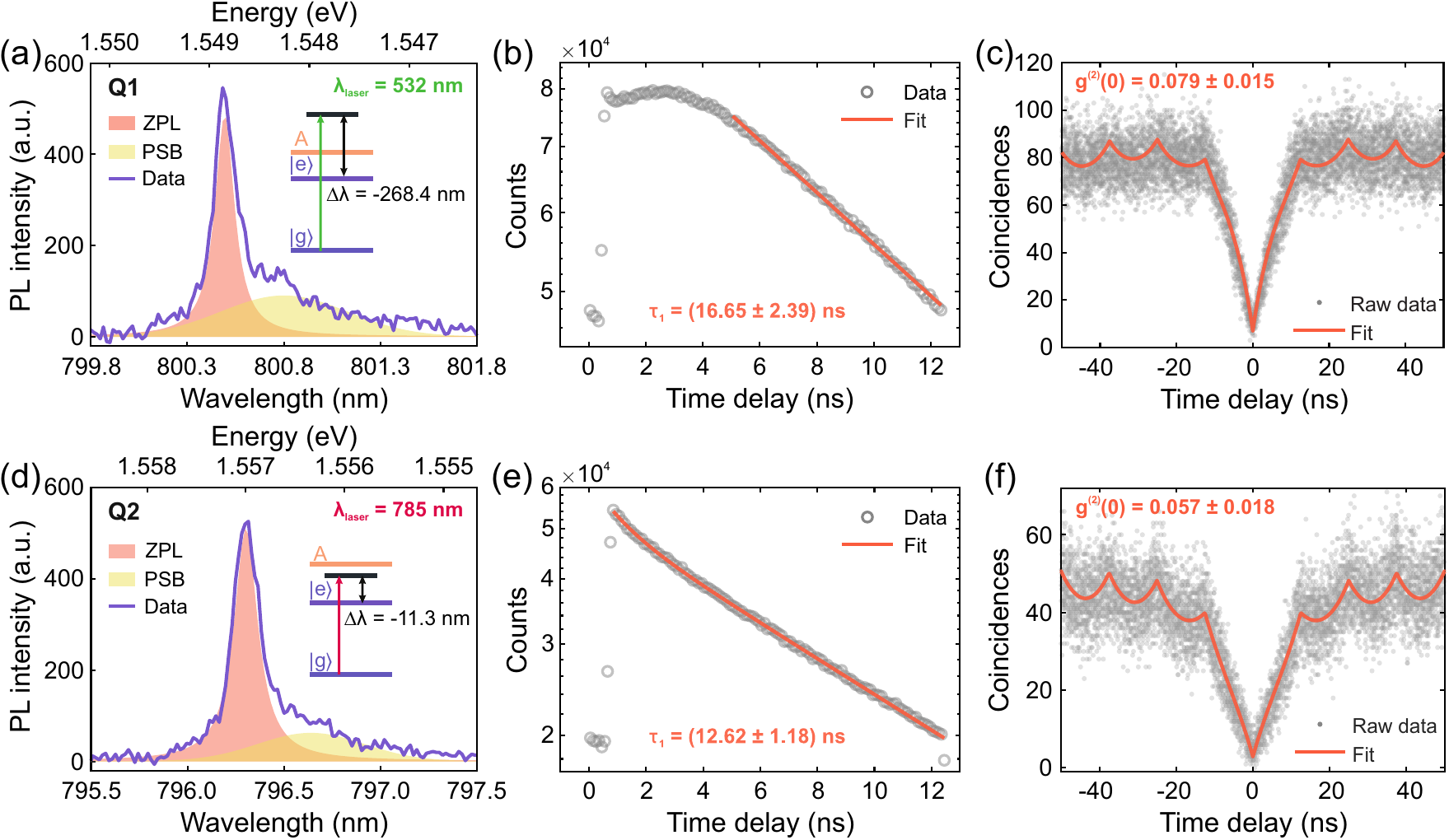}
\caption{ \textbf{Above-band and near-resonant excitations.} Emission spectra of emitter Q1 under pulsed above-band excitation at 532 nm \textbf{(a)} and emitter Q2 under near-resonant pulsed excitation at 785 nm \textbf{(d)}. The graphs present fittings to the raw data (solid purple line) with a sum of a Lorentzian (red area) and a Gaussian (yellow area) accounting for the contribution from the ZPL and the PSB, respectively. The inset schemes on the right show a simplified band diagram of the emitters as a two-level system ($\ket{g}$ and $\ket{e}$) with the respective pumping energy, while A represents the energy level of the planar WSe\textsubscript{2} A-exciton. \textbf{(b)} Semi-logarithmic plot of the time-resolved PL measurement from Q1 (grey circles) with relative single decay exponential fitting function (red line) with a constant $\tau_1 = (16.65 \pm 2.39)$ ns. \textbf{(c)} Second-order intensity correlation measurement ($g^{(2)}(\tau)$) of Q1 under above-band pulsed excitation. From the double exponential fit (red line), we extract an antibunching value of $g^{(2)}(0) = 0.079\pm 0.015$. \textbf{(e)} Semi-logarithmic plot of the time-resolved PL measurement from Q2 (grey circles) with relative double decay exponential fitting function (red line). The two extracted time constants are $\tau_{1} = (12.62 \pm 1.18)$ ns and  $\tau_2 = (1.14 \pm 0.21)$ ns, accounting for 96.5\% and 3.5\% of the fit curve, respectively. \textbf{(f)}  Second-order intensity correlation measurement of Q2 under near-resonant excitation. The $g^{(2)}(0)$ value extracted is $0.057 \pm 0.018$. The measurements in (b) and (e) are integrated over 3 minutes, while those in (c) and (f) over 8 hours.}
\label{Figure2}
\end{figure*}

We measured the single-photon purity of bilayer WSe$_2$ emitters by measuring the second-order correlation of a representative emission line (Q1) under CW excitation at \SI{470}{\nano\meter}. A \SI{750}{\nano\meter} long-pass filter helps suppress the emitted light from the laser light. The measurement is performed with a Hanbury Brown-Twiss (HBT) setup, which includes a 50:50 fiber beam splitter and a pair of superconducting nanowire single-photon detectors (SNSPDs).
In Fig.~\ref {Figure1}(c), we show the obtained correlation histogram as a function of delay, which exhibits a strong antibunching behavior. The suppressed coincidence counts at zero time delay establish the clear single-photon nature of the emission with extracted single-photon purity of $g^{(2)}(0) = 0.098 \pm 0.045$.

\section{Above-band vs. near-resonant optical excitation}
\label{section:above-bandVSnear-resonant}
To systematically compare the effect of the above-band and the near-resonant excitation schemes on the bilayer quantum emitters, we conducted an investigation of the optical properties of two emitters (Q1 and Q2). Our study includes high-resolution emission spectra, lifetime, and purity. We analyze the differences to find the most effective approach for optimal performance.

\vspace{10pt}


\textbf{Above-band optical excitation.}
Here, we used a \SI{532}{\nano\meter} pulsed laser excitation. Fig. \ref {Figure2}(a) shows the \textmu PL spectrum obtained from Q1 at \SI{4}{\kelvin} (see Fig. S4 of the Supplementary Information for the temperature-dependent series up to \SI{40}{\kelvin}), revealing a narrow zero-phonon line (ZPL) at \SI{800.5}{\nano\meter} and a visible broad phonon side-band (PSB). By fitting a Lorentzian and a Gaussian function to the ZPL and to the PSB, respectively, we qualitatively extract a width of \SI{0.108}{\nano\meter} (\SI{0.209}{\milli\electronvolt}) for the ZPL and \SI{0.782}{\nano\meter} (\SI{1.513}{\milli\electronvolt}) for the PSB, and a Debye-Waller factor $ \mathrm{DWF} = {I_{\mathrm{ZPL}}}/({I_{\mathrm{ZPL}}+I_{\mathrm{PSB}}}) $ of 0.551 (Fig. \ref{Figure2}(a)). 
This suggests a strong exciton-phonon coupling, which is in good agreement with previous reports  \cite{Vannucci2024Single-photonIndistinguishability, Schneider2017Exciton-polaritonEngineering, Reitzenstein2011ElectricallySources}.
Notably, the ZPL is orders of magnitude broader than its Fourier-limited value of \SI{57.0e-6}{\nano\metre} (\SI{0.11}{\micro\eV}), considering the decay time of \SI{16.65}{\nano\second} (see next paragraph). This suggests that the pure dephasing and charge noise contributions to the total linewidth is substantial, consequently resulting in lower indistinguishability \cite{Akbari2022Lifetime-LimitedModulation}.
Additionally, we observe that the width of the emission line gets broader when increasing the laser excitation power, almost doubling its value at the saturation level, as shown in Fig. S2 of the Supplementary Information.

For the characterization of the dynamic behavior of an individual quantum emitter under above-band excitation, we carried out a time-resolved PL (TRPL) measurement. As shown in Fig. \ref{Figure2}(b), we observe a relatively fast decay at shorter time delays ($\leq $ \SI{0.5}{\nano\second}), after which the curve rises again and reaches its maximum at around \SI{3}{\nano\second}. We attribute this to trapped states involved in the population of the excited state \cite{Bacher1999BiexcitonDot, Dass2019Ultra-LongHeterostructures}.
At longer delay times, the decay is fitted with a single exponential (red line), revealing a decay time of $\tau_1\! =\! (16.65 \pm 2.39)\,$ns (uncertainty given as standard error). Such value is in good agreement with previous observations on TMD single-photon sources \cite{Amani2016RecombinationDichalcogenides, Ye2017Single2, Paralikis2024TailoringEngineering}.
The second-order correlation measurement under pulsed above-band excitation, shown in Fig. \ref{Figure2}(c), reveals a $g^{(2)}(0)$ of $0.079 \pm 0.015$. The extracted value is consistent with the one in Fig. \ref {Figure1}(c) under CW operation, indicating a good single-photon purity also under triggered excitation.


\vspace{10pt}

\textbf{Near-resonant optical excitation.} 
To perform near-resonant excitation, the wavelength of the pulsed laser was varied from \SI{730}{\nano\meter} to \SI{790}{\nano\meter}, corresponding to a detuning $\Delta\lambda = \lambda_{\rm laser} - \lambda_X$ from \SI{-66.3}{\nano\meter} to \SI{-6.3}{\nano\meter} with respect to the quantum emitter ($\lambda_X = $ \SI{796.3}{\nano\meter}), in steps of \SI{10}{\nano\meter}. The energy corresponding to this pumping wavelength is below the energy of the A-exciton of planar WSe\textsubscript{2} \cite{Terrones2014BilayersHeterostructures}, but above the excitonic transition energy considered ($\ket{g} \rightarrow \ket{e}$) (cf. the band diagram scheme in the inset of Fig. \ref{Figure2}(d)), thus, we refer to such a detuning as \textit{near-resonant}. 
We observed that the quantum emitter could be excited over a large range of detunings (see Supplementary Information Note S3 for more details). This behavior has been previously reported in $\mathrm{WSe}_{2}$ \cite{Tonndorf2015Single-photonSemiconductor, vonHelversen2023Monolayer} based on photoluminescence excitation spectroscopy measurement.
In our case, the maximum emission intensity occurred when the laser was detuned $\Delta\lambda$ = \SI{-11.3}{\nano\meter} (\SI{9.1}{\milli\electronvolt}) from the emission line, with results reported in Figs. \ref{Figure2}(d)-(f). 
At this detuning, the zero-phonon line and phonon sideband linewidths corresponding to emitter Q2 are found to be \SI{0.124}{\nano\meter} and \SI{0.760}{\nano\meter}, respectively. From this we compute the Debye-Waller factor as $\mathrm{DWF} = 0.538$. (Fig.~\ref{Figure2}(d)). Additionally, under near-resonant excitation, we achieved a $g^{(2)}(0)$ value of $0.057 \pm 0.018$ (Fig.~\ref{Figure2}(f)). These results are similar to those obtained under the above-band excitation and suggest that the emission spectrum and single-photon purity of these particular quantum emitters may not be heavily dependent on the excitation scheme.

However, a striking dissimilarity between the two excitation schemes arises from the time-resolved measurement. As shown in Fig. \ref{Figure2}(e), the lifetime histogram obtained under near-resonant excitation exhibits a predominant single exponential decay contribution with a time constant of $\tau_1\!=\!(12.62 \pm 1.18)\,$ns. This particular contrast in decay dynamics under near-resonant excitation as compared to the above-band measurement presented in Fig.~\ref{Figure2}(b) can be attributed to the quantum emitter being optically excited near its bandgap. These experimental results suggest that directly populating a bilayer quantum emitter with a proper excitation scheme is essential to avoid unnecessary internal relaxation processes and the creation of additional unwanted photo-induced charge carriers, resulting in simpler dynamics and reduced lifetime.


\begin{figure*}[hbt!]
\includegraphics[width=0.95\linewidth]{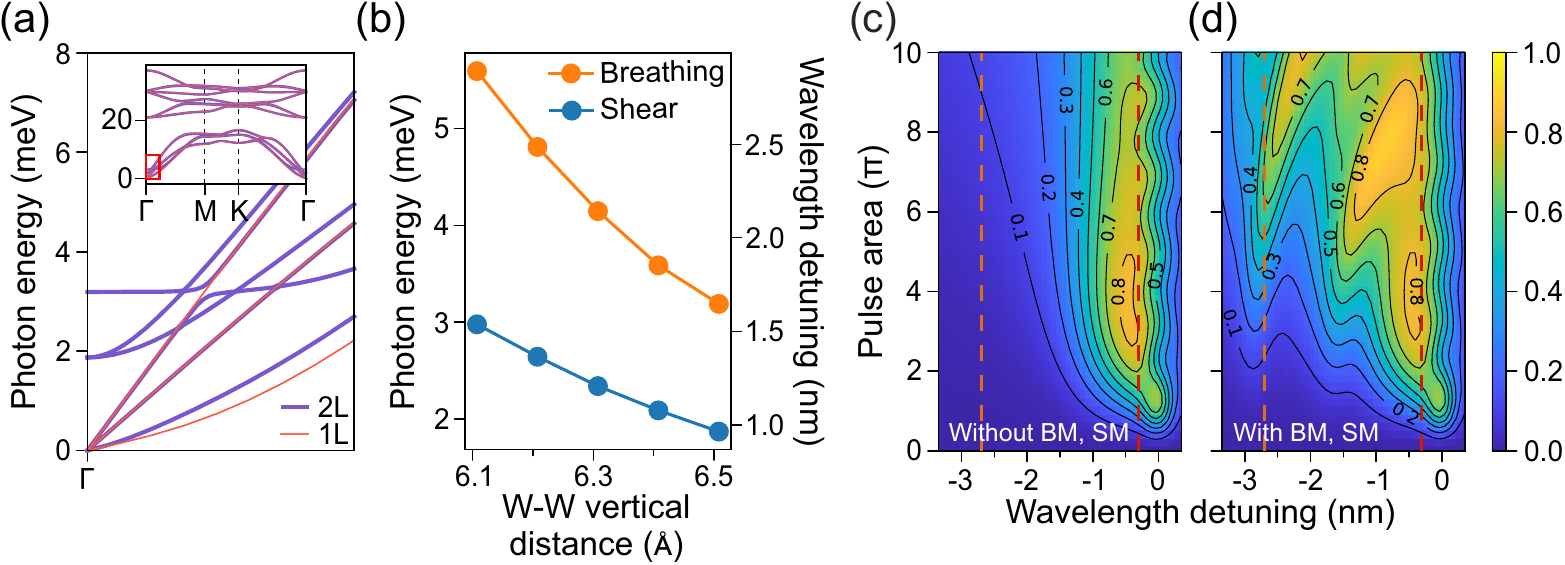}
\caption{\textbf{DFT-based phonon signatures on the population inversion for mono- and bilayer WSe$_{2}$.}
\textbf{(a)} Calculated phonon dispersion in mono- and bilayer WSe$_{2}$ around the $\Gamma$ point. For the bilayer, the equilibrium value $d_{\rm W-W} = 6.5$ Å is used for the vertical distance between W atoms in different layers. The inset shows the phonon dispersion across the full Brillouin zone. 
\textbf{(b)} Energy of interlayer SMs and BM at $\Gamma$ as a function of the interlayer W-W distance $d_{\rm W-W}$.
\textbf{(c)} Calculated population inversion in the presence of LA phonon coupling, as a function of wavelength detuning $\lambda_{\rm laser} - \lambda_X$ from the exciton and pulse area $\Theta$. The laser pulse has a FWHM of 0.30 nm ($t_p$ = 2.65 ps in our notation). \textbf{(d)} Same as in (c), including also the coupling to SMs and BM. Here, we use $d_{\rm W-W} = 6.1$ Å for the interlayer W-W distance, and the dimensionless BM coupling weight is set to $\xi_{\rm BM} = 5$ (see \textit{Methods}). Dashed orange and red lines are placed at detunings of \SI{-2.7}{nm} and \SI{-0.3}{nm}, respectively.
}
\label{Figure3}
\end{figure*}

\rev{
\section{Phonon-assisted excitation}}
\label{section:phonons}

\rev{
\subsection{Theory: signatures of exciton-phonon coupling in the population inversion of bilayer WSe$_{2}$ QEs}}
\label{section:theory_phonons}

\rev{
We now consider a laser excitation closer to resonance mediated by phonon-assisted relaxation, with a detuning of only a few nanometers or sub-nanometer. To support our experimental results, we first develop a theory of phonon-coupled QEs in bilayer WSe$_2$, and calculate the population inversion under pulsed laser excitation. 
Earlier experimental and theoretical reports have shown evidence of strong exciton-phonon coupling in WSe$_2$.
While low-energy phonon coupling in monolayer WSe$_2$ is mainly due to 2D LA phonons~\cite{Mitryakhin2024Electrodynamics, vonHelversen2023Monolayer, Vannucci2024Single-photonIndistinguishability}, bilayer structures have a richer phonon landscape including interlayer vibrational modes~\cite{Jin2016InterlayerHeterostructures, Ripin2023TunablePhononicCoupling}.

First, the phonon bandstructure of pristine mono- and bilayer WSe$_2$ is calculated with Density Functional Theory (DFT) and shown in Fig.~\ref{Figure3}(a), see \textit{Methods} for details. In the low-energy sector around $\Gamma$, monolayer WSe$_2$ supports three phonon branches with vanishing energy at $\Gamma$, corresponding to out-of-plane, transverse, and longitudinal modes respectively.
In contrast, bilayer WSe$_2$ has two additional shear modes (SMs), involving lateral sliding between layers, and one breathing mode (BM), characterized by the out-of-plane compression and expansion of layers. Their energies at the $\Gamma$ point of the Brillouin zone are \SI{1.9}{meV} and \SI{3.2}{meV}, respectively.
Both SMs and the BM are inherently absent in monolayers.

Importantly, SMs and BM are sensitive to the interlayer distance, which is likely modified in the strain-engineered nanowrinkle. As shown in Fig.~\ref{Figure3}(b), the energy of SM and BM phonon modes increases to \SI{3.0}{meV} and \SI{5.6}{meV}, respectively, when the distance $d_{\rm W-W}$ between W atomic planes is compressed from the equilibrium value of \SI{6.5}{\AA} to \SI{6.1}{\AA}.
This indicates that the optical signature of SM (BM) phonons moves from a detuning of \SI{-0.97}{nm} (\SI{-1.65}{nm}) with respect to the emission line to \SI{-1.54}{nm} (\SI{-2.89}{nm}) under a small relative compression of 6\% in the out-of-plane direction.
Larger compression shifts the SM and BM further towards higher energy.

Next, we calculate the population inversion under pulsed laser excitation for emitter Q1 and for wavelength detuning up to $\Delta \lambda$ = \SIlist{-3.4}{\nano\meter}.
We calculate the excited state population after a Gaussian electric field pulse $\Omega(t) = \frac{\Theta}{\sqrt{\pi} t_p} e^{-(t / t_p)^2}$, with $\Theta = \int_{-\infty}^{+\infty} \dd{t} \Omega(t)$ the pulse area. The model includes the influence of the phonon environment with a numerically exact process tensor formalism (see \textit{Methods}).
The calculation includes a suitable spectral broadening to reproduce the experimental spectrum.
In Fig.~\ref{Figure3}(c), we consider only the coupling to LA phonons, which occurs in both mono- and bilayer structures.
A sizeable population $> 0.8$ is observed at detuning $\Delta \lambda \approx -0.4$ nm and for a moderate pulse area of 3--5$\pi$ (for the population inversion results at different laser parameters, see Supplementary Information Note S6).
The signal decreases monotonically for larger detuning.
We interpret these results as a signature of LA phonon-assisted excitation, where a laser pulse with a shorter wavelength than the exciton is able to populate the excited state by exciting additional phonon modes. The optimal detuning corresponds indeed to the maximum of the LA phonon spectral density.
Closer to resonance ($|\Delta \lambda| < $ \SI{0.3}{nm}), we observe the signature of Rabi oscillations as a function of the pulse area. We interpret these as resonant excitation of the emitter with the tail of the laser pulse, whose spectral width is FWHM = \SI{0.30}{nm}. For narrower laser pulses, oscillations are confined to the strictly resonant condition $\Delta \lambda \approx 0$ (see Supplementary Note S6).

The inclusion of SM and BM phonon coupling changes the scenario drastically. As seen in Fig.~\ref{Figure3}(d), the exciton population after the laser pulse evolves non-monotonically as a function of the wavelength detuning. For $\Theta < 5 \pi$, a first peak is observed at $\Delta \lambda \approx -2.8$ nm and corresponds to phonon-assisted pumping via the BM. A second peak occurs around $\Delta \lambda \approx -1.5$ nm and is caused by coupling to SM phonons. Finally, LA phonon-assisted excitation emerges at a shorter detuning. For larger pulse areas, the signature of BM and SM phonons shifts closer to the exciton and tends to merge with the LA band.

It should be noted that the spectral density of SM and BM coupling has been calculated under the assumption that the emitter-phonon coupling mechanism is the same as for LA modes, however, the contribution of the BM coupling in Fig.~\ref{Figure3}(d) has been magnified by a factor $\xi_{\rm BM} = 5$ to obtain qualitative agreement with the subsequent experiment. 
Optical signatures of BM phonons in the population inversion are also visible at $\xi_{\rm BM} = 1$, as reported in Supplementary Note S6 and Figure S7.
A detailed study of the BM and SM coupling mechanism deserves further investigation in a separate work.
}


\begin{figure}[t!]
\includegraphics[width=\columnwidth]{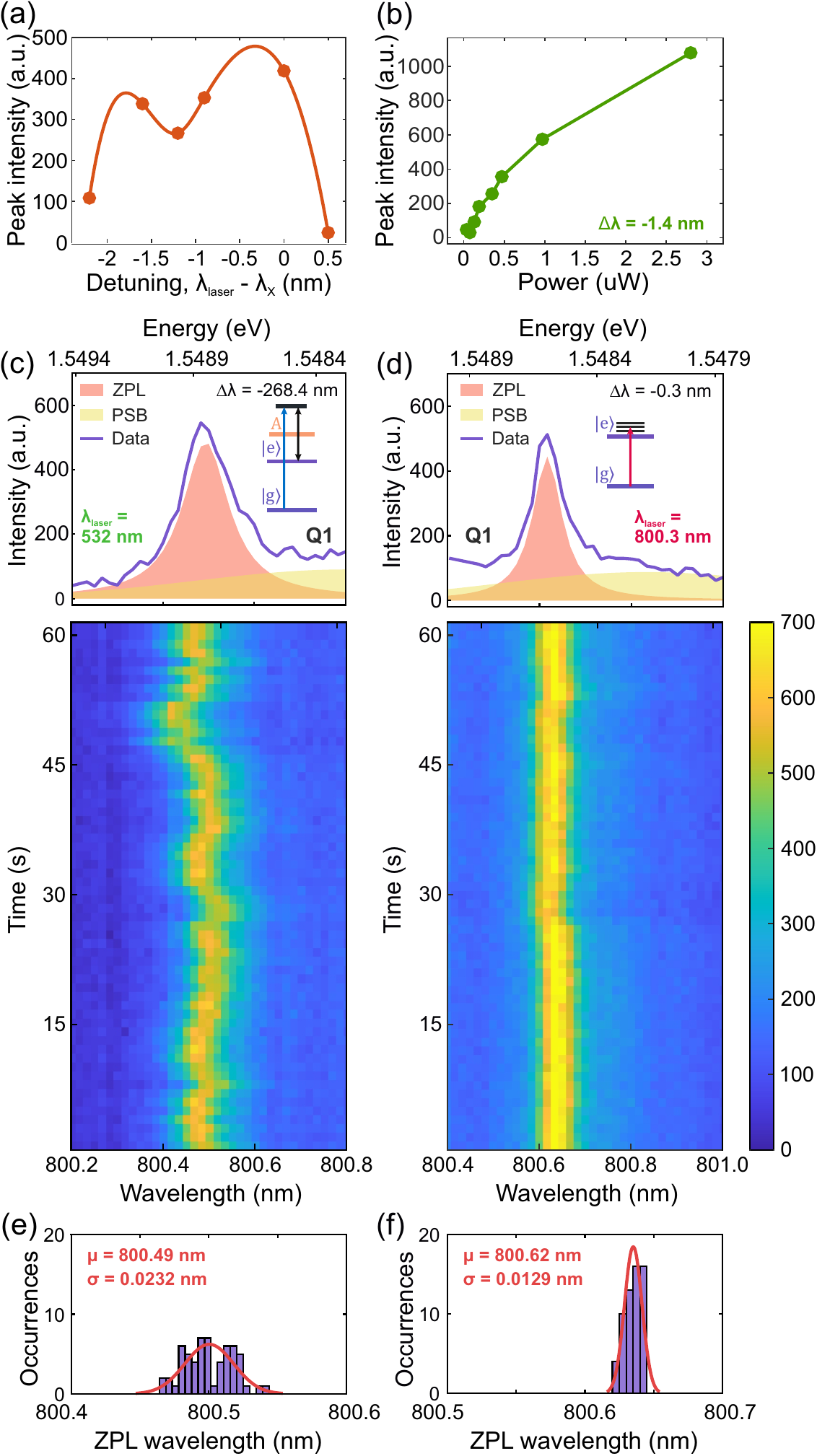}
\caption{
{\rev{\textbf{Longitudinal acoustic phonon-assisted excitation and spectral diffusion.} \textbf{(a)} Coarse PLE spectrum of the quantum emitter, showing two absorption maxima centered at detuning $\Delta \lambda\!=\,$\SI{-1.7}{\nano\meter} and $\Delta \lambda\!=\,$\SI{-0.3}{\nano\meter}, respectively. \textbf{(b)} Power dependence of the peak intensity evaluated at a detuning $\Delta \lambda\!=\,$\SI{-1.4}{nm}.}} High-resolution PL spectra at time \SI{1}{\second} \textbf{(c top, d top)} and their time evolution \textbf{(c bottom, d bottom)} for above-band excitation and phonon-assisted excitation, respectively. \textbf{(e)} and \textbf{(f)} show the statistical analysis of the central wavelength of the ZPL as extracted from the fitting of each line in the map. The red curves are Gaussian fittings of the obtained distributions. The mean value and the standard deviation in (e) are \textmu \textsubscript{ZPL} = 800.49 nm and \textsigma \textsubscript{ZPL} = 0.0232 nm, respectively. At the same time, their values for (f) are \textmu \textsubscript{ZPL} = 800.62 nm and \textsigma \textsubscript{ZPL} = 0.0129 nm.}
\label{Figure4}
\end{figure}


\rev{
\subsection{Longitudinal acoustic modes and spectral diffusion.}}
\label{section:longitudinal_phonons}

\rev{
To experimentally demonstrate phonon-assisted excitation on bilayer WSe$_2$ quantum emitters, first, we identified a single isolated emission line ($\lambda\!=\,$\SI{808.2}{nm}) under the above-band excitation of a 532 nm pulsed laser. Then we performed a photoluminescence excitation (PLE) measurement with smaller laser detuning to the emission line scanning a constant power pulsed laser, ranging from  $\Delta \lambda\!=\,$\SI{-2.2}{\nano\meter} to $\Delta \lambda\!=\,$\SI{+0.5}{\nano\meter} (Fig.~\ref{Figure4}(a)). The obtained coarse PLE measurement reveals two maxima. Fitting the data points, the two maxima appear to be centered at $\Delta \lambda\!=\,$\SI{-1.7}{\nano\meter} and $\Delta \lambda\!=\,$\SI{-0.3}{\nano\meter} respectively, in qualitative agreement with the theoretical prediction for SM and LA modes. The signal goes down to negligibly small values at positive detuning, after crossing the ZPL. 
The power-series measurement in Fig.~\ref{Figure4}(b) at $\Delta \lambda\!=\,$\SI{-1.4}{\nano\meter} detuning shows a saturation behavior of the peak emission intensity, which could be interpreted as a signature of exciton preparation through incoherent processes \cite{Glassl2013ProposedRobustPhonons, Thomas2021BrightDipole}. While carrying out the PLE measurement close to the resonance condition, the emission line under LA phonon-assisted excitation clearly appears to be more stable and narrower in contrast to the emitter excited under the above-band excitation.

To quantitatively evaluate and compare the line broadening and the spectral diffusion under different excitation schemes, we performed \textmu PL time evolution measurements from the same emitter Q1 under phonon-assisted and above-band excitation. For the phonon-assisted measurements, the excitation laser detuning was fixed at $\Delta \lambda\!=\,$\SI{-0.3}{\nano\meter}.
On the other hand, for the above-band excitation measurement, a 532 nm pulsed laser ($\Delta \lambda\!=\,$\SI{-268.4}{\nano\meter}) was used to excite the emitter Q1, as in Fig.~\ref{Figure2}(a,b,c).
The emission line excited via the acoustic phonons (Fig.~\ref{Figure4}(c) top) exhibits a FWHM of the fitted \textmu PL spectrum of \SI{0.074}{\nano\meter} (\SI{151}{\micro\eV}). This value is close to the nominal resolution of our spectrometer (\SI{0.05}{\nano\meter}), which might, thus, not allow us to resolve the true linewidth of the emission. Anyhow, the obtained value is 1.5 times narrower than the FWHM of \SI{0.108}{\nano\meter} from the spectrum under the below-saturation above-band excitation  (Fig.~\ref{Figure4}(d) top).

The broadening of the peak's linewidth provides an indication of energy shift of the quantum emitter energy in timescales faster than the experimental acquisition time, which is one of the main impediments to indistinguishable photon generation. On the other hand, spectral diffusion of the order of 1Hz can be traced over the measurement timescale. In this regard, we then recorded the emission signal as a function of time under both excitation schemes, as shown in the bottom panel of Figs.~\ref{Figure4}(c) and (d).
Remarkably, the spectral diffusion at a timescale slower than the experimental acquisition time is also drastically reduced under acoustic phonon-assisted excitation. In both maps, each horizontal frame corresponds to 1 second of measurement, during which the PL emission is integrated over \SI{0.7}{\second}. By fitting each spectrum with a Lorentzian function, we can extract the resonance of the ZPL as a function of time. The histogram plots in Figs. \ref{Figure4}(e) and (f) show the number of occurrences in the wavelength range of emission, grouped up in \SIlist{0.005}{\nano\meter}-broad bins. Fitting the histograms with a Gaussian function allows us to extract a mean value of \textmu \textsubscript{ZPL}$\,=\,$\SIlist{800.49}{\nano\meter} and a standard deviation of \textsigma \textsubscript{ZPL}$\,=\,$\SIlist{0.0232}{\nano\meter} for the distribution of the peaks' central position under above-band excitation, with \textmu \textsubscript{ZPL}$\,=\,$\SIlist{800.62}{\nano\meter} and \textsigma \textsubscript{ZPL}$\,=\,$\SIlist{0.0129}{\nano\meter} under acoustic phonon-assisted excitation.
}

This suggests that a more stable charge environment is achieved under phonon-assisted excitation thanks to the reduced creation of additional photo-induced charges compared to above-band excitation, resulting in lower spectral diffusion.
Both the inhomogeneous broadening of the ZPL and its spectral diffusion, here mitigated, are critical obstacles to single-photon indistinguishability, which requires a lifetime-limited linewidth, identical emission energy, and well-defined polarization.
Improved reduction of inhomogeneous broadening and spectral wandering of the emission peak could be achieved by combining the current pulsed excitation scheme with a more beneficial substrate material choice \cite{Iff2017SubstrateEngineeringArchitectures}.

\rev{
Despite the efficacy of excitation via acoustic phonons, the laser leakage into the ZPL of the emission line under such a small detuning deteriorates the single-photon purity.
Filtering out the laser leakage is a practical challenge that arises when the detuning is in the order of the sub-nanometer. As a consequence, in the following, we aim to perform excitation through BM absorption, as suggested from Fig.~\ref{Figure3}(d). As mentioned before, the BM mode is a unique feature of bilayer structures, and its appearance at higher spectral detuning from the ZPL compared to the LA modes makes it more practically feasible to perform efficient excitation of bilayer QEs for high-purity single-photon generation with uncomplicated spectral filtering to suppress the pump laser. 
}

\vspace{10pt}

\subsection{Breathing mode and single-photon characterization.}
\label{section:breathing_mode}


\begin{figure*}[hbt!]
\includegraphics[width=0.75 \textwidth]{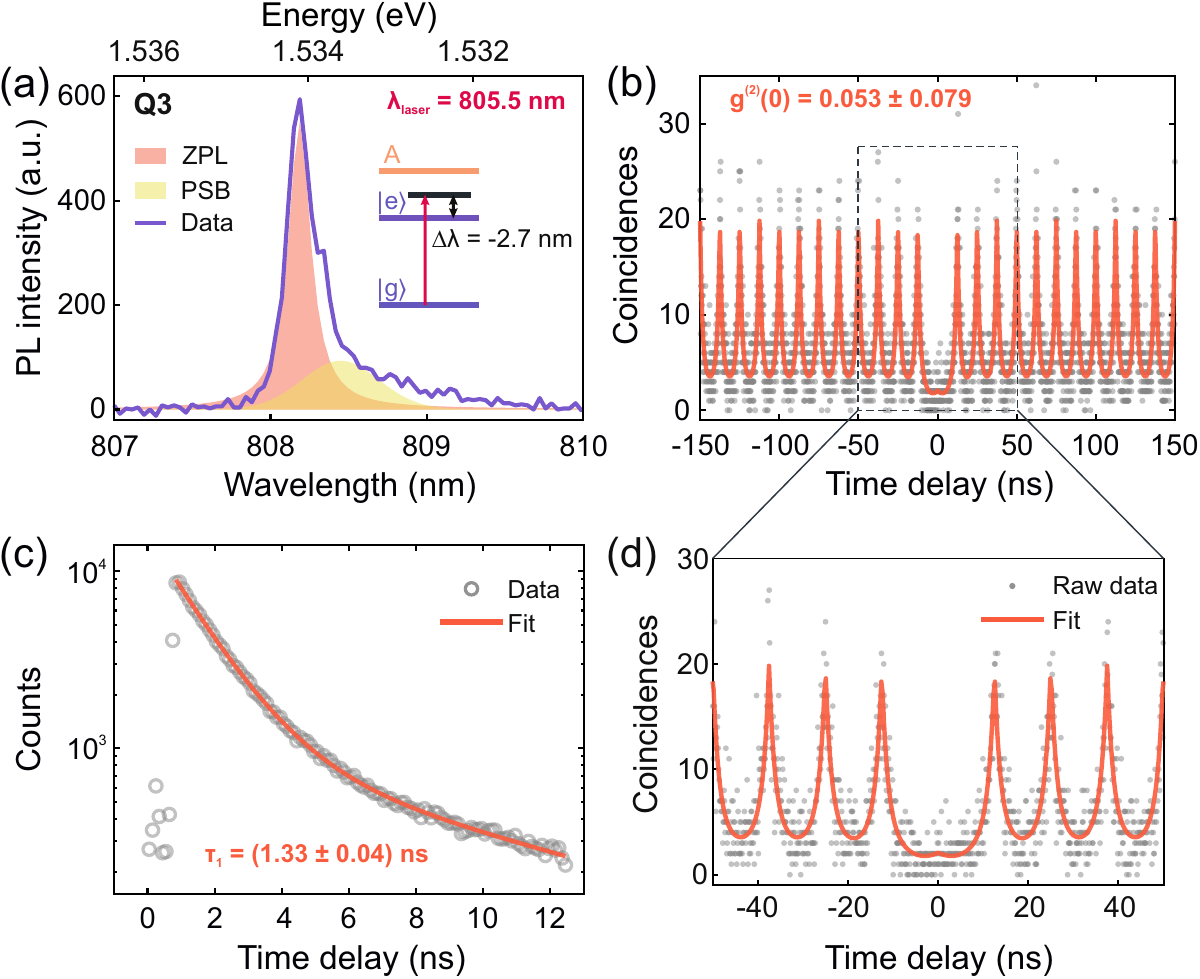}
\caption{\textbf{Phonon-assisted excitation.} \textbf{(a)} PL spectrum of emitter Q3 under pulsed excitation at detuning of \SIlist{-2.7}{\nano\meter} from the quantum emitter. The fit function on the raw data (purple) is a sum of a Lorentzian (red area) and a Gaussian (yellow area) accounting for the contribution from the ZPL and the PSB, respectively. The inset shows a simplified band diagram of an emitter as a two-level system highlighting the pumping energy detuning. \textbf{(b)} Second-order autocorrelation measurement of Q3. From the double exponential fit (red line), we extract an antibunching value of $g^{(2)}(0) = 0.053 \pm 0.079$. \textbf{(c)} Time-resolved PL measurement from Q3 in a semilogarithmic plot with relative double exponential fitting function with extracted time constants of $\tau_{1} = (1.33 \pm 0.04)$ ns (80.4\%) and  $\tau_{2} = (8.31 \pm 6.29)$ ns (19.6\%). \textbf{(d)} Zoom-in of $g^{(2)}(\tau)$ in (c) for time delays up to $\pm 50$ ns.}
\label{Figure5}
\end{figure*}

\rev{
To exploit the BM phonon for the exciton preparation of bilayer quantum emitters, we tuned the laser at wavelength $\lambda_{\rm laser}\!=$ \SIlist{805.5}{\nano\meter}, which corresponds to a detuning of $\Delta\lambda\!\approx$ \SIlist{2.7}{\nano\meter} (\SIlist{5}{\milli\eV}) from the Q3 line, centered at \SIlist{808.2}{\nano\meter} (Fig.~\ref{Figure5}(a)).
By coarsely scanning the laser wavelength,  we experimentally observed that the emission intensity is rather sensitive to laser spectral shift, completely disappearing for detunings deviating $\pm$ \SIlist{0.2}{nm} from the optimal condition of $\Delta\lambda\! =$ \SIlist{-2.7}{\nano\meter}.
It is worth mentioning that such larger laser detuning (-2.7 nm) of the BM excitation compared to the excitation via LA phonons makes spectral filtering of the laser light from the single-photon emission more straightforward.}
The extracted FWHM with a Lorentzian fit is \SI{0.174}{\nano\meter} (\SI{0.33}{\milli\electronvolt}), approximately 40\% larger compared to Q1 and Q2 under above-band and near-resonant excitation. Note that the Debye-Waller factor under this excitation has increased to 0.712.
The broader ZPL compared to those in Figs.~\ref{Figure2}(a) and (d) under above-band and near-resonant excitations, respectively, can be attributed to the fact that pulsed excitation at a detuning of \SIlist{-2.7}{\nano\meter} (\SIlist{5}{\milli\eV}) requires high excitation power, potentially resulting in additional line broadening.
The second-order autocorrelation measurement (see Fig.~\ref{Figure5}(b)) shows a $g^{(2)}(0)$ of $0.053\ \pm\ 0.079$, establishing highly pure single-photon emission similar to those under above-band and near-resonant excitations.
\rev{This testifies to the success of using the BM to perform excitation of the bilayer quantum emitter and obtain high single photon purity thanks to easier spectral filtering.}

However, the most remarkable observation is the shortened decay time of the emission under phonon-assisted excitation. The double exponential fit on the time-resolved PL data presented in Fig. \ref{Figure5}(c) suggests that 80.4\% of the contribution to the emitted light has a decay time of $(1.33 \pm 0.04)$ ns. \rev{This is five times shorter than the minimum decay time recorded under above-band excitation in mono- and bilayer WSe$_2$ from the flake at issue, and among the shortest decay times reported in the literature of WSe\textsubscript{2}-based quantum emitters on a dielectric substrate without any cavity (see Supplementary Information Note S7 and Fig. S10). Therefore, the lifetime measurement under phonon-assisted excitation can confidently be attributed to the employed excitation mechanism rather than to variability among different quantum emitters.}
We attribute this reduction of about one order of magnitude as compared to the above-band excitation to the fact that exciting near the resonance of the emitter via the efficient acoustic phonons avoids additional relaxation processes before populating the system.
The relevance of this result lies in the fact that faster exciton recombination paves the way for a deterministically triggered source at a higher rate for future quantum computing devices based on WSe\textsubscript{2} quantum emitters. Besides that, the additional low contribution (below 20\% of the total integrated counts) of longer decay time $\tau_2 = (8.31 \pm 6.29)$ ns can be attributed to additional slow carrier recombination processes within the bilayer quantum emitter.
\rev{A previous report on monolayer WSe$_2$ quantum emitter shows CW excitation at a detuning of \SI{5.07}{meV} \cite{Kumar2016ResonantWSe_2}, corresponding to \SI{-2.5}{nm}, where coupling to a blue-shifted exciton (BS-X) resonance is identified. The energy detuning is very similar to the energy of the breathing mode here considered. However, the work in Ref.~\cite{Kumar2016ResonantWSe_2} is fully conducted on monolayer WSe$_2$, which does not support interlayer SM and BM phonons.
Even though we cannot rule out the possibility of absorption from an additional charged or dark exciton state, the predictions from our theoretical model in the presence of SM and BM coupling are consistent with our experimental observation.
}


\section{Discussion}
\label{section:discussion}

Our results demonstrate that the exciton preparation scheme substantially affects the characteristics of the emitted single photons. Employing a proper pumping strategy is necessary to achieve high purity, fast dynamics, and minimal spectral diffusion. All the excitation schemes presented in this work rely on incoherent processes. In particular, the phonon-assisted scheme provides efficient population inversion and improved spectral stability, inescapable premises for a future efficient and indistinguishable source of single photons from bilayer WSe\textsubscript{2}.
\rev{Our theoretical model of the exciton preparation, including the QE coupling to LA phonon modes, SMs, and BM, is in qualitative agreement with the experiments and provides a pathway towards optimization of the excitation strategy.
However, the exact physical mechanism behind the SM and BM coupling is not fully understood and deserves further investigation.
Moreover, the coupling to longitudinal optical (LO) phonon modes, higher-order dark states~\cite{Lindlau2018TheWSe2} and charged states~\cite{Kumar2016ResonantWSe_2} is necessary for an accurate and realistic representation of the physical platform under study, eventually providing a deeper understanding of the exciton preparation and recombination mechanisms.}

Yet, coherent state preparation is essential to increase the coherence time and to generate superposition states \cite{Senellart2017High-performanceSources}, which are base requirements for quantum computing and quantum information processing. In this regard, the strictly resonant (in a cross-polarization configuration) and the Swing-UP of the quantum EmitteR population (SUPER) excitation \cite{Bracht2021Swing-upPulses, Karli2022SUPEREmitter, Boos2022CoherentTechnique, Vannucci2024Single-photonIndistinguishability} were pursued in this work.
Preliminary investigations reveal that resonant excitation is achieved even though the high linear polarization of the emitter dipole does not allow for the complete rejection of the back-scattered laser (see Supplementary Note S8), making it challenging to proceed with more detailed optical characterization. On the other hand, no emission was observed under the SUPER excitation. The main obstacles encountered could be related to low- and high-frequency noise in the charge environment, as well as strictly practical complications, such as the rejection of the high laser power necessary to perform this scheme, which required both spectral and cross-polarization filtering.

Nonetheless, the outcomes from the acoustic phonon-assisted excitation measurements, combined with the knowledge acquired during the coherent and SUPER excitation attempts, suggest that charge stabilization is imperative for the advancement of the WSe\textsubscript{2} platform. This can be achieved by electrical control via the implementation of external electrical biasing. Having contacted devices will also provide control of the in-plane and out-of-plane electric fields, allowing spectral tuning of the emission by Stark shift.
On the other hand, increasing the efficiency of Wse$_2$ quantum emitters by enhancing the spontaneous emission rate can be achieved through the implementation of the emitter in an optical cavity \cite{Flatten2018MicrocavityEnhancedWSe2, Iff2021Purcell-EnhancedCavity}.
Additionally, understanding the exciton fine structure splitting in this type of quantum emitters will enable its control \cite{Glazov2022ExcitonFineMonolayers} and the development of tailored coherent excitation strategies.

\rev{
Despite the incoherent nature of phonon-assisted excitation scheme~\cite{Steinhoff2025impactphononlifetimessinglephoton}, its exploitation in terms of exciton preparation fidelity, purity, and indistinguishability of the emitted photons is sufficient to demonstrate six-photon boson sampling and heralded three-photon entanglement with semiconductor QDs \cite{Maring2024AVersatilePlatform}.
Moreover, it offers improved stability and robustness against laser detunings and power levels,} as well as the possibility of exciting and collecting the photons in all polarization directions.


\section{Conclusion}
\label{section:conclusion}

\rev{
We have systematically investigated the single-photon properties of bilayer WSe\textsubscript{2} quantum emitters under various optical excitation schemes, including above-band, near-resonant, and phonon-assisted excitation via longitudinal acoustic (LA) and breathing phonon (BM) modes. Our combined experimental and theoretical approach provides new insights into exciton-phonon coupling and its role in efficient exciton preparation in the case of bilayer WSe\textsubscript{2} quantum emitters.
Our theoretical calculations predict an efficient population above 80\% under LA phonon-assisted excitation at a detuning of \SI{-0.4}{nm}. Additionally, we anticipate a sizable population at larger detunings of  \SI{-2.7}{nm}, dependent on the strength of exciton-BM coupling. Experimentally, we demonstrated that phonon-assisted excitation via the BM mode resulted in a faster recombination time of ($1.33 \pm 0.04$) ns - more than an order of magnitude shorter than above-band and near-resonant excitations - while maintaining a high single-photon purity of 0.947$\,\pm\,$0.079. The consistently high purity across all excitation schemes suggests that the emission purity is not strongly excitation-dependent. However, LA phonon-assisted excitation achieved a near resolution-limited zero-phonon line width of 0.072 nm and a 1.8-fold reduction in spectral diffusion compared to above-band excitation, demonstrating its advantage in emitter stability.
These findings highlight the critical role of the excitation scheme in achieving high-fidelity single-photon emission from bilayer WSe\textsubscript{2}. In particular, phonon-assisted excitation offers a promising pathway to minimizing spectral diffusion and improving emitter stability, representing a significant step toward realizing transform-limited linewidths and enhancing the indistinguishability of emitted photons—key requirements for scalable quantum technologies.
Understanding the non-trivial correlation between excitation schemes and single-photon emission characteristics provides valuable insights for the development of deterministic, high-quality quantum light sources based on the TMD platform.
}

\section{Methods}
\label{section:methods}
\noindent \textbf{Sample fabrication and preparation:} The sample consists of a bilayer WSe\textsubscript{2} transferred on top of a 3-point star-shaped SiO\textsubscript{2} (HSQ) nanopillar on a Si/SiO\textsubscript{2} substrate.
For the pillar fabrication, the sample was spin-coated with a negative electron-beam resist (Fox-06), and the star-shaped structures were patterned using a \SI{100}{\kilo\volt} electron beam lithography machine (JEOL 9500) with a dose of \SI{11000}{\micro\coulomb\per{\centi\meter\squared}} and an electric current of \SI{6}{\nano\ampere}. The exposed resist was then developed in a (1:3) solution of AZ400K:H\textsubscript{2}O. The HSQ resist is known to chemically change to SiO\textsubscript{2} after e-beam exposure, resulting in SiO\textsubscript{2} nanostructures of the same height as the resist ($\approx$ \SI{150}{\nano\meter}).
The WSe\textsubscript{2} flake was exfoliated from the bulk crystals using the conventional scotch-tape technique and subsequently transferred to the patterned substrate by deterministic dry-transfer procedure.
Finally, defects were introduced in the flake lattice by electron-beam irradiation of the strained areas using the same \SI{100}{\kilo\volt} EBL machine with a dose of \SI{1000}{\micro\coulomb\per{\centi\meter\squared}} and a current of \SI{6}{\nano\ampere}. \\

\noindent \textbf{Optical characterization:} To perform photoluminescence measurements, we placed the WSe\textsubscript{2} mono and bilayer samples in a closed-cycle helium cryostat (attoDRY 800, attocube systems AG) with a base temperature of \SI{4}{\kelvin}. The cryostat is equipped with a microscope objective to focus the laser light onto the sample (NA = 0.8, $60\times$,  attocube systems AG). The setup is configured in a cross-polarized configuration. For photoluminescence imaging, we employ a cooled sCMOS camera (Dhyana 400D, 4MP, Tucsen).
To enable various excitation schemes, we employed a tunable OPO module (Chameleon Compact OPO-Vis, Coherent Inc.) pumped by a Ti:Sapphire laser (Chameleon Ultra II, Coherent Inc.) producing \SI{150}{\femto\second} pulses at an \SI{80}{\mega\hertz} repetition rate. To further optimize the laser spectral window for various excitation methods, we also employed a home-built 4$f$ pulse shaper ($f$$\,=\,$\SI{500}{\milli\meter}) equipped with an \SI{1800}{groove\per\milli\meter} diffraction grating and a programmable single-mask transmissive spatial light modulator (SLM-S320, Jenoptik) in the Fourier plane. The SLM consisting of 320 pixels, helps to carve a narrow spectral range (FWHM $\approx$ \SI{0.21}{\nano\meter}). 
To acquire the emission spectra, we utilized a fiber-coupled spectrometer (Horiba iHR 550) using either a \SI{600}{groove\per\milli\meter} or a \SI{1200}{groove\per\milli\meter} grating. For second-order correlation measurements, we employed a fiber-based Hanbury Brown and Twiss (HBT) interferometer coupled to superconducting nanowire single-photon detectors (SNSPD, ID218, ID Quantique) and time taggers (ID900, ID Quantique).

\textbf{Data analysis:} All the PL emissions from quantum emitters presented in the manuscript were fitted with a function that is a sum of a Lorentzian and a Gaussian. The Lorentzian contribution accounts for the homogeneous broadening of the line, while the Gaussian distribution fits the phonon sideband. From the fitting, it is possible to extract the complete set of 6 parameters defining the distributions, together with the standard error for each one of them.
The data obtained from time-resolved photoluminescence (TRPL) measurements was fitted with a sum of two uncorrelated exponentially decaying functions added to a constant background: $ F_{\rm TRPL}(\tau) = F_0 + A_{1} e^{-(\tau-\tau_{01}) / \tau_1} \, + \, A_{2} e^{-(\tau-\tau_{02}) / \tau_2}$, where $F_0$ accounts for the constant background, $A_{1}$ and $A_{2}$ are the amplitudes of the two exponential functions, $\tau_1$ and $\tau_2$ the two extracted decay constants, and $\tau_{01}$ and $\tau_{02}$ the decay offsets. The relative weights of the two are extracted as a percentage. In this way, we gain insight into the types of transitions that will contribute to the purity measurement.
For the analysis of the second-order correlation measurements, the raw data is fitted with the following expression: $ F_{\rm SOC}(\tau) = B_{1} [e^{-|\tau| / \tau_1} \, + \, e^{-|\tau| / \tau_2} ] \, + \, B_{2} \sum_{n} [e^{-|\tau+n\tau_{\text{rep}}| / \tau_1} + e^{-|\tau+n\tau_{\text{rep}}| / \tau_2} ] $, where $B_{1}$ denotes the central peak area, $B_{2}$ the side peak areas, $\tau_1$ and $\tau_2$ the two extracted decay time constants, $\tau_{\text{rep}}$ the laser pulse repetition time, and $n$ the index of each peak ($n=0,\pm1,\pm2, ...$). The $g^{(2)}(0)$ value is then calculated as $g^{(2)}(0) = F_{\rm SOC}(0)/\max[F_{\rm SOC}(\tau)]$.

\rev{
\textbf{Theory---DFT calculations:}
Ab-initio DFT calculations were performed within the Atomic Simulation Recipes (ASR) Python framework \cite{Gjerding2021ASR}, using the electronic structure code GPAW \cite{Mortensen2024GPAW} and the Atomic Simulation Environment (ASE) library \cite{Larsen2017ASE}, and  managed with the MyQueue job
scheduler \cite{Mortensen2020MyQueue}.
Monolayer and bilayer atomic structures were obtained from the C2DB \cite{Gjerding2021RecentC2DB} and BiDB \cite{Pakdel2024BiDB} databases, respectively. 
We used a plane wave basis set with a cutoff energy of
800 eV, uniform k-point grid of density 6.0/Å$^{-1}$, and Fermi-Dirac smearing of 50 meV.
The Perdew-Burke-Ernzerhof (PBE) \cite{PBE1996} exchange functional was used, including the DFT-D3 \cite{Grimme2010DFTD3} dispersion correction to account for van der Waals interlayer interactions.
Atomic forces for phonon calculations were converged to 10$^{-4}$ eV/Å. The phonon dispersion was calculated with the Phonopy package \cite{Togo2023PhonopyJPCM, Togo2023PhonopyJPSJ}.
}

\rev{
\textbf{Theory---Open quantum system dynamics}:
The predicted population inversion under near-resonant excitation is obtained from the theoretical modeling of a QE coupled to phonons and to a Gaussian laser pulse, and assuming that electron-phonon interaction occurs via the deformation potential mechanism.
We consider a two-level system (ground state $\ket{G}$, excited state $\ket{X}$) with Hamiltonian $H_0 = \hbar \omega_X^{(0)} \dyad{X}$ coupled to a bath $H_{\rm ph} = \sum_m \sum_{\vb k} \hbar \omega_{m, \vb k} b^{\dag}_{m, \vb k} b_{m, \vb k}$ via the interaction $H_{\rm int} = \sum_m \sum_{\vb k} \hbar g_{m, \vb k} \qty( b^\dag_{m, \vb k} + b_{m, \vb k}) \dyad{X}$.
Here, the label $m$ runs over different phonon modes (LA, SM$_1$, SM$_2$, and BM).

The influence of the linear LA mode (dispersion $\omega_{{\rm LA}, \vb k} = c |\vb k|$, with $c = 4.4 \times 10^3$ m s$^{-1}$ obtained by fitting the DFT result) is summarized by the spectral density
\begin{equation}
    J_{\rm LA}(\omega) = \alpha \omega^2 \exp \qty(-\frac{\omega^2}{\omega_{\rm c}^2}) ,
\end{equation}
with $\alpha = (D_e - D_h)^2 / (4 \pi \hbar \rho c^4)$ = 0.29 ps obtained from first principles as outlined in Ref.~\cite{Vannucci2024Single-photonIndistinguishability}. Here, $D_e$ ($D_h$) is the deformation potential for electrons (holes), and $\rho$ the area density of a single layer of WSe$_2$. The cutoff frequency is determined as $\omega_{\rm c} = 2.03$ THz from the measured LA sideband.

For the two SM modes (SM$_1$ and SM$_2$, with degenerate energy at $\Gamma$), we fit the DFT phonon dispersion to the relativistic-like formula $\omega_{{\rm SM}_i, \vb k} = \sqrt{c_i^2 |\vb k|^2 + M^2}$, which is approximately linear for $|\vb k| \gg M / c_i$ but has nonzero frequency $M$ at the $\Gamma$ point. We obtain the spectral density
\begin{equation}
    J_{{\rm SM}_i}(\omega) = \alpha \frac{\omega^2 - M^2}{r_i^4} \exp \qty(-\frac{\omega^2 - M^2}{r_i^2 \omega_{\rm c}^2}) ,
\end{equation}
with $r_i = c_i / c$, see details in Supplementary Note S5.

The BM is modeled as a dispersion-less mode with frequency $\omega_{\rm BM}$ obtained from the DFT value at $\Gamma$, resulting in a delta-like spectral density
$J(\omega) = \alpha \omega_{\rm c}^4 (2 \omega_{\rm BM})^{-1} \delta(\omega - \omega_{\rm BM})$.
To circumvent the numerical singularity, we use instead a Gaussian spectral density centered around $\omega_{\rm BM}$,
\begin{equation}
    J_{\rm BM}(\omega) = \xi_{\rm BM} \frac{\alpha \omega_{\rm c}^4}{2 \omega_{\rm BM}} \frac{1}{\sqrt{\pi} \varepsilon} \exp \qty[-\qty(\frac{\omega - \omega_{\rm BM}}{\varepsilon})^2],
\end{equation}
with width $\varepsilon=0.2$ THz accounting for a finite phonon lifetime. Additionally, we introduce a dimensionless weight $\xi_{\rm BM}$ to control the strength of the BM coupling.
}

To model the laser excitation, the system is also coupled to a pulsed laser at frequency $\omega_{\rm laser}$ via $H_{\rm laser} = \frac{\hbar}{2} \Omega(t) \qty(e^{-i \omega_{\rm laser} t} \ketbra{X}{G} + e^{i \omega_{\rm laser} t} \ketbra{G}{X})$, with a Gaussian pulse $\Omega(t) = \frac{\Theta}{\sqrt{\pi} t_p} e^{-(t / t_p)^2}$ and $\Theta = \int_{-\infty}^{+\infty} \dd{t} \Omega(t)$ the pulse area.
The dynamics is obtained by solving the master equation for the reduced density operator $\rho(t)$ with the formalism of time-evolving matrix product operators (TEMPO) \cite{Strathearn2018EfficientOperators} as implemented in the \textsc{OQuPy} Python package \cite{Fux2024OQuPyAPythonPackage}, allowing to treat the phonon environment at a numerically-exact level.
\rev{
We use four separate process tensors, one per each phonon mode.
From the phonon spectral density, we calculate the mode-dependent Huang-Rhys factor as $S_m = \int_0^{+\infty} \dd{\omega} \omega^{-2} J_m(\omega)$, and the total Huang-Rhys factor as $S = \sum_m S_m$.
To account for spectral broadening beyond the transform limited value, we add to the master equation a Lindblad term $\gamma \mathcal L_{\sigma^\dag \sigma}[\rho]$, where $\mathcal L_A[\rho] = A \rho A^\dag - \frac 1 2 \acomm{A^\dag A}{\rho}$.
}

To calculate the $\ket{X}$ state population $P_X$ after the laser pulse, the system is initialized in $\ket{G}$ at time $t=-3 t_p$ and $P_X$ is calculated as $P_X = \Tr \qty[\dyad{X} \rho(t)] \eval_{t = 3 t_p}$.
\rev{The pulse detuning in frequency is defined as $\Delta \omega = \omega_{\rm laser} - \omega_X$, where $\omega_X = \omega_X^{(0)} - D$ is the exciton frequency shifted by the total polaron shift $D = \sum_m D_m$, and $D_m = \int_0^{+\infty} \dd{\omega} \omega^{-1} J_m(\omega)$.}
The corresponding wavelength detuning is $\Delta \lambda = \lambda_{\rm laser} - \lambda_X = 2\pi c \qty(\omega_{\rm laser}^{-1} - \omega_X^{-1})$.
\rev{For a given value of $t_p$, the pulse FWHM in wavelength is calculated from the squared amplitude of the Fourier-transformed signal, $\qty|\widehat \Omega(\omega)|^2 = \frac{\Theta^2}{2 \pi} \exp \qty(-\frac 1 2 t_p^2 \omega^2)$, and is given by $\Delta \lambda_{\rm FWHM} = \lambda_X^2 \sqrt{2 \ln 2} / (\pi c t_p)$.}

\section*{Associated Content}
\noindent
\textbf{Supplementary Information}

\noindent
The Supplementary Information is available free of charge at \url{https://...................}

\section*{Notes}
The authors declare no competing financial interest.

\section*{Acknowledgement}
The authors acknowledge support from the European Research Council (ERC-StG ``TuneTMD", grant no. 101076437), and the Villum Foundation (grant no. VIL53033). The authors also acknowledge the European Research Council (ERC-CoG ``Unity", grant no. 865230), Carlsberg Foundation (grant no. CF21-0496), and support from the Independent Research Fund Denmark (Grant DFF-9041-00046B). NG acknowledges support from the European Union's Horizon 2020 Research and Innovation Program under the Marie Skłodowska-Curie Grant Agreement no. 861097. VR acknowledges financial support through the Austrian Science Fund FWF projects 10.55776/TAI556 (DarkEneT) and  10.55776/FG5. The authors acknowledge the cleanroom facilities at DTU Nanolab – National Centre for Nano Fabrication and Characterization.

\section*{Author contributions}

AP fabricated the final samples with support from CP and AM. CP performed all the optical measurements from the fabricated samples. CP and VR assembled a home-built pulse-shaping setup using the SLM. CP, PW, and AP contributed to experimental data analysis and processing. LV performed DFT calculations, modeled the coupling to shear and breathing phonon modes, and calculated the exciton population. LV and JN modeled the coupling to longitudinal acoustic phonons and fitted the experimental data. BM conceived the idea, and LV, NG, and BM coordinated the project. CP took the lead in writing the manuscript. All authors discussed the results and contributed to the final manuscript.

\section{References}
\bibliographystyle{naturemag}
\bibliography{references}

\end{document}